\documentclass[amssymb,prb,aps,twocolumn,showpacs,10pt]{revtex4}
\usepackage{graphics,graphicx}

\begin{document}
\title{Dynamics of Macroscopic Tunneling in Elongated BEC}
\author {G. Dekel$^1$}
\author{V. Farberovich$^1$}
\author{V. Fleurov$^1$}
\author{A. Soffer$^2$}
\affiliation{$^1$ Raymond and Beverly Sackler Faculty of Exact
Sciences, School of Physics and Astronomy, Tel-Aviv University,
Tel-Aviv 69978 Israel}
\affiliation{$^2$ Department of Mathematics, Rutgers University, New
Brunswick, NJ 08903,USA}

\begin{abstract}
We investigate macroscopic tunneling from an elongated quasi 1-d
trap, forming a 'cigar shaped' BEC. Using recently developed
formalism\cite{ours1} we get the leading analytical approximation
for the right hand side of the potential wall, i.e. outside the
trap, and a formalism based on Wigner functions, for the left side
of the potential wall, i.e. inside the BEC. We then present
accomplished results of numerical calculations, which show a 'blip'
in the particle density traveling with an asymptotic shock velocity,
as resulted from previous works on a dot-like trap, but with
significant differences from the latter. Inside the BEC a pattern of
a traveling dispersive shock wave is revealed. In the attractive
case, we find trains of bright solitons frozen near the boundary.
\end{abstract}
\pacs{82.20.Xr, 03.75.Kk, 05.90.+m} \maketitle

\section{Introduction}

In recent years, nonlinear dynamics of BEC out of equilibrium has
been given a great amount of attention, mainly in looking for
various nonlinear coherent structures such as dark, bright and
oblique solitons, vortices and dispersive shock waves, and studying
their emergence and evolution. Dark and bright solitons were first
observed experimentally in Refs.
\onlinecite{darkso1,darkso2,brightso1,brightso2}. Vortices in 2d
BECs were discussed in Refs.
\onlinecite{vortex1,vortex2,vortex3,vortex4}. Recently, much
attention has been drawn to dispersive shock waves, which contrary
to their dissipative counterparts in compressible fluid, and due to
their quantum nature are controlled by the dispersion effects rather
than by dissipation, and are now believed to emerge in and dictate
the dynamics of BEC flow. They are characterized by an expanding
oscillatory front, and, as other phenomena in BEC, are predicted and
quested by the the Gross - Pitaevskii equation (GPE)
\begin{equation}\label{GPEbec}
i \hbar\frac{\partial\Psi(r,t)}{\partial t} =
$$$$
\left[-\frac{\hbar^2}{2m} \nabla^2 + U_{ext}(r) + NU_{int} \mid
\Psi(r,t) \mid^2\right]\Psi(r,t)
\end{equation}

Recently dispersive shock waves in different BEC setups and with
various initial conditions have been investigated and predicted
theoretically, as well as observed in experiments. First evidence of
possible shock waves development in BEC setups were reported in
\cite{shock1}, where a sharp density depression was induced by slow
light technique. Theoretical studies showed formation of a shock
front in traveling 1-d BEC wave packets, split from an initial
density perturbation in Refs. \onlinecite{shock3a} and
\onlinecite{shock3b}. Shock waves in BEC induced by using Feshbach
resonance were studied in Ref. \onlinecite{shock4} and Witman
averaging method was used to analyze 1-d BEC shock waves in the
small dispersion limit in Ref. \onlinecite{shock5}. Ref.
\onlinecite{shock2} presented imaging of rotating BEC "blast waves"
along with numerical analysis. A broad comparison between dispersive
and dissipative shock waves in all dimensions was carried out in
Ref. \onlinecite{shock6}, which included examination of experimental
reports and theoretical results. Study of supersonic flow past a
macroscopic obstacle in 2-D BEC gave rise to predictions of a sonic
Cherenkov cone that would eventually transform into spatial 2-d
supersonic dispersive shock waves.\cite{shock7} Predictions of new
2-d creatures, called oblique solitons followed and are now an
object of extensive studies in the field.\cite{shock8,shock9}. In
Refs. \onlinecite{fleischer1,fleischer2}, evidences of shock
phenomena in macroscopic flow were found in nonlinear optics
experiments.

In our previous works\cite{ours1,ours2} we studied tunneling from a
trapped BEC Gaussian packet. For this purpose we analytically solved
GPE in its hydrodynamic presentation. We showed that tunneling
resulted in formation of an isolated 'blip' in the particle density
outside the trap, originating from a shock-type solution of the
Burgers equation and moving with the asymptotic velocity of this
shock. Experimental studies of the relevant phenomena in nonlinear
optics were reported in Refs. \onlinecite{bpss08,bpsss08} This
effect is independent of the effect of non-linearity, i.e.
inter-atomic interaction in BEC or Kerr nonlinearity optics, and
includes the case in which the latter is zero.\cite{ours1}. This
effect can be made repeatable in time by periodically bringing the
potential walls closer (or changing the trap frequency) thus
releasing additional blips with feasible control of their
parameters, which may lead to an eventual realization of atom
soliton laser.\cite{ours2} Since a strongly trapped BEC packet,(i.e.
the margins of the trap are of the same order of magnitude as its
bulk in all directions) reacts to tunneling in pulsations inside the
trap, with frequency which is twice the in-trap eigen frequency,
this periodicity is used to determine the timing of the mechanism
described above. Sudden turn-on of a matter-wave source in order to
create repeated pulses was discussed in Ref. \onlinecite{CMM07}.

This paper will investigate tunneling from a quasi 1d cigar shaped
trap with a twofold motivation: Firstly, we intend to look for the
emergence of a single blip and its properties in order to eventually
create a pulsed atom laser on the basis of this system, providing a
much greater source of matter than in the previous case of tunneling
from a trapped narrow packet (tightly trapped in all directions).
Our second aim is to investigate the opposite effect, i.e. the
dynamics inside the BEC cigar shaped trap, which appears to be much
richer. We will show that an "anti-blip" (a local depletion) is
formed, which propagate inside the trap with the same velocity as
the blip outside but in the opposite direction. However the
structure and development of this anti-blip is very complicated and
in particular it may develop into a dispersive shock wave.

\section{Model and Formalism}

\begin{figure}[tbp!]
\begin{center}
\includegraphics[width=8cm]{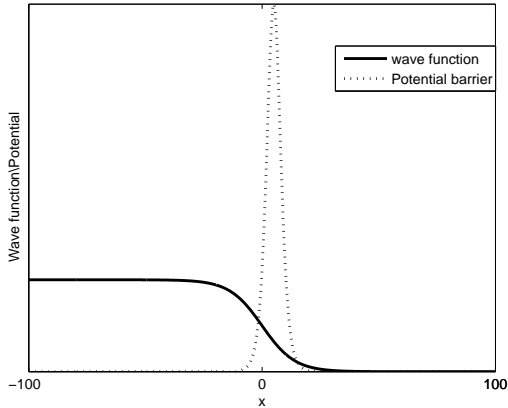}
\caption{A schematic arrangement of the investigated problem. At
$t=0$, the potential wall narrows down to a finite width barrier,
enabling the cigar shaped BEC tail to tunnel}\label{cigbarr}
\end{center}
\end{figure}

We consider the following physical system. An elongated 1d
wave-function is sealed on its right hand side by a high enough
wall, i.e. $U_0 > NU_{int}\mid \psi(r,t) \mid^2$. (The quantum
pressure term is always negative around the barrier region) The wall
is lowered, at $t=0$, so that a tail of the wave function is allowed
to tunnel and dynamically evolve according to the GPE through the
new barrier of a finite width, the height of its peak is that of the
former wall. This is shown in Fig. \ref{cigbarr}. We solve the 1-d
GPE (\ref{GPEbec}) numerically and analytically for the right and
left sides of the barrier respectively, after $t=0$.

\section{Analytical results}

\subsection{Right hand side of the barrier}

In our previous work\cite{ours1} we introduced a formalism that
solves GPE with a given external potential, in the hydrodynamic
representation. The GPE is equivalently written in the form of the
Euler type
\begin{equation}\label{euler}
v_t + \frac{1}{2} \nabla v^2
=-\frac{1}{m}\left(-\frac{\hbar^2\nabla^2 \sqrt{\rho}}
{2m\sqrt{\rho}}+U_{ext}+NU_{int}\right)
\end{equation}
and continuity
\begin{equation}\label{continuity}
\rho_t + \nabla(\rho v)=0.
\end{equation}
equations. Then the Euler equation (\ref{euler}) is solved under an
adiabatic approximation according to which the density field $\rho =
|\Psi|^2$ evolves with the time of tunneling $\tau_{tun}$ whereas
the velocity field $m{\bf v} = \nabla\varphi$ ($\varphi$ being the
gradient of the phase of wave function $\Psi$) with the traversal
time $\tau_{tr}$ satisfying the condition $\tau_{tun}\gg \tau_{tr}$.
The problem of a quantum fluid dynamics can be mapped onto that of a
classical dissipationless fluid motion. We found out that solving
the first iteration actually sufficed to investigate short and long
time dynamics of the problem.

\begin{figure}[tbp!]
\begin{center}
\includegraphics[width=7cm]{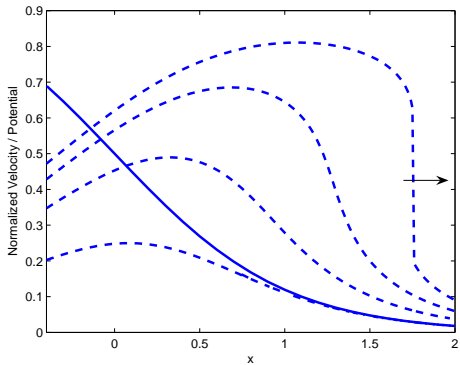}
\caption{(Color online) Velocity field distribution at the first
stages of tunneling from the cigar shaped trap. A tendency to shock
formation is clearly observed. At $t=4$ there is a steep gradient (a
jump) which indicates a transition to triple solution region
}\label{velocitycigar}
\end{center}
\end{figure}

To investigate the dynamics outside the barrier in the present work,
the analysis can be therefore made, as in the previous
works\cite{ours1} by solving the classical integral
\begin{equation}\label{time}
t = \int_{x_0}^x \frac{d\xi}{\sqrt{\frac{2}{m}(\epsilon - U(\xi))}}
\end{equation}
where
\begin{equation}\label{potential}
U(\xi) = U_{t=0^-}(\xi)- U_{t=0^+}(\xi) = \frac{1}{2}U_0(1 -
\tanh(\alpha \xi)
\end{equation}
is the difference between the potential shapes before and after
switching from a wall to a barrier of finite width. $t$ is the time
required for a fluid droplet (tracer) to reach the point $x$ and
have velocity $v$, if it has started from the point $x_0$ with the
energy $\epsilon$. Assuming that initially the fluid was motionless
$ v_0=0$  at $t=0$, i.e. $\epsilon = mv^2/2$, we get
\\
\begin{widetext}
\begin{equation}\label{timeb}
\alpha t(x,\epsilon) = \frac{\mbox{arctanh}( \sqrt{\frac{ \epsilon -
\frac{1}{2} U_0 + \frac{1}{2} U_0 \tanh(\alpha x)}{\epsilon -
U_0}})}{\sqrt{\epsilon - U_0}} + \frac{\mbox{arctanh}(
\sqrt{\frac{\epsilon - \frac{1}{2} U_0 + \frac{1}{2} U_0
\tanh(\alpha x)}{\epsilon}})}{\sqrt{\epsilon}}
\end{equation}
\end{widetext}
>From here, one can find $\epsilon(x,t)$ numerically and therefore
the velocity field $v(x,t)$. The results are shown in Fig.
\ref{velocitycigar} and indicate, as expected, a tendency to shock
wave formation (compare Refs. \onlinecite{ours1,ours2}) to result in
a blip in the density distribution. The time scale of its creation
is $4$ time units (traversal times $\tau_{tr}$) which matches that
of the blip formation time to be obtained numerically.

\subsection{Left hand side of the barrier}\label{left}

To deal with the dynamics to the left side of the barrier (inside
the BEC) analytically, one can derive a hydrodynamic "hole
representation" by means of Wigner function technique. For this, let
us assume that we have a homogeneous density distribution $\rho_0$
and consider fluctuations changing this density. In this case we
have to define a "hole Wigner function" as
\begin{equation}\label{Wigner-a}
f^h_W({\bf r},{\bf p}, t) =
$$$$
\int \frac{d^3y}{(2\pi\hbar)^3} \left[ - \psi({\bf x} - {\bf y}/2,t)
\psi^*({\bf x} + {\bf y}/2,t) + \rho_0\right] e^{i{\bf py}/\hbar} =
$$$$
- f_W({\bf r}, {\bf p}, t) + \rho_0 \delta({\bf p}).
\end{equation}
We define the hole density  by integrating (\ref{Wigner-a}) over the
momentum {\bf p}
\begin{equation}\label{holedens}
\rho_h({\bf r}, t) = \int d^3p f^h_W({\bf r},{\bf p}, t) = -
\rho({\bf r}, t) + \rho_0
\end{equation}
The hole velocity field is defined as
\begin{equation}\label{holevel} m_0
\rho_h({\bf r}, t) v^\alpha_h({\bf r}, t) = \int d^3p\ p^\alpha
f^h_W({\bf r},{\bf p}, t) =
$$$$
\rho_h({\bf r}, t) \overline{p_h^\alpha} = - \rho({\bf r}, t)
\overline{p^\alpha}
\end{equation}
Using definitions (\ref{holevel}) and (\ref{holedens}) we may write
the density continuity equation in the form
\begin{equation}\label{continuity-hole}
\frac{\partial}{\partial t} \rho_h({\bf r}, t) + \mbox{\boldmath
$\nabla$} \cdot (\rho_h({\bf r}, t){\bf v}_h({\bf r}, t)) = 0.
\end{equation}

In order to get the momentum continuity equation we first
introduce the second conditional moments in the hole
representation
$$
\rho_h({\bf r}, t) \overline{p_h^\alpha p_h^\beta} = \int d^3p\
p^\alpha p^\beta\ f^h_W({\bf r}, {\bf p}, t) =
$$$$
\int d^3p\ p^\alpha p^\beta\ f_W({\bf r}, {\bf p}, t) = - \rho({\bf
r}, t) \overline{p^\alpha p^\beta}
$$
and use the Liouville-Moyal equation
\begin{equation}\label{L-M}
\frac{\partial}{\partial t} f_W({\bf r},{\bf p}, t) =
$$$$
\frac{1}{i\hbar}\{H({\bf r},{\bf p}) \star f_W({\bf r},{\bf p}, t)
- f_W({\bf r},{\bf p}, t) \star H({\bf r},{\bf p})\}
\end{equation}
which governs the dynamics of Wigner function,\cite{m49} where
$$
H({\bf r},{\bf p}) = \frac{p^2}{2m} + V({\bf r})
$$
is the classical Hamiltonian of the system. The definition of the
$\star$ product and detailed discussion can be found in Ref.
\onlinecite{Z02}. Multiplying equation (\ref{L-M}) by $p^\alpha$ and
integrating over $p$ we get
\begin{equation}\label{euler1-hole}
- m_0 \frac{\partial}{\partial t} [\rho_h({\bf r}, t)
v^\alpha_h({\bf r}, t)] =
$$$$
- \frac{\partial V({\bf x})}{\partial x^\beta} \rho({\bf r}, t)
\delta^{\alpha\beta} + \frac{1}{m_0} \frac{\partial }{\partial
x^\beta} [\rho_h({\bf r}, t) \overline{p_h^\alpha\ p_h^\beta}]
\end{equation}

Then using the continuity equation (\ref{continuity-hole}) we write
\begin{equation}\label{euler2-hole}
- m_0 \rho_h({\bf r}, t) \frac{\partial v_h^\alpha({\bf r},
t)}{\partial t} - m_0 \rho_h({\bf r}, t) \left[v_h^\beta({\bf r}, t)
\frac{\partial }{\partial x^\beta}\right] v_h^\alpha ({\bf r}, t) =
$$$$
- \frac{\partial V({\bf x})}{\partial x^\alpha} \rho_h({\bf r}, t)
+ \frac{1}{m_0} \frac{\partial }{\partial x^\beta}
\left\{\rho_h({\bf r}, t) \left[ \overline{p_h^\alpha\ p_h^\beta}
- \overline{p_h^\alpha}\ \overline{ p_h^\beta}\right]\right\}
\end{equation}
Till now no approximations have been made. But now we need an
approximation that the holes can be described by the single wave
function $\psi_h({\bf r},t) = f_h e^{-\frac{i}{\hbar} S_h}$ which
can be justified for small values of the hole density $\rho_h({\bf
r},t) << \rho_0$. Then the standard procedure (see, e.g. Ref.
\onlinecite{lf01}) results in
\begin{equation}\label{euler3}
m_0 \frac{\partial {\bf v}_h({\bf r}, t)}{\partial t} + m_0
\left[{\bf v}_h({\bf r}, t) \cdot \mbox{\boldmath $\nabla$}\right]
{\bf v}_h({\bf r}, t) =
$$$$
- \mbox{\boldmath $\nabla$} \left\{-\frac{\hbar^2}{2
m_0}\frac{\mbox{\boldmath $\nabla$}^2 f_h}{f_h} -  V({\bf
x})\right\}
\end{equation}
which is the Euler equation for the holes similar to that for
particles analyzed in Refs.\onlinecite{ours1,ours2}. That is why in
order to study the dynamics of tunneling inside the cigar shaped
trap we may use the same technique as described in Ref.
\onlinecite{ours1}. Due to the symmetry of the problem we may
directly apply equations (\ref{time}) - (\ref{timeb}) resulting in
an emission of a blip in the empty space right of the barrier and
come to the conclusions that simultaneously a depletion (an
anti-blip) should be formed to the left of the barrier. It is
expected to propagate to the left with the same velocity as the blip
propagates to the right of the barrier.

However it is clear that the symmetry is not complete. In particular
the density of particles to the right of the barrier cannot become
negative. On the other hand, the density to the left may locally
become higher than $\rho_0$, i.e. "negative density of holes" is
possible. Hence a more complicated behavior of the "anti-blip" is
expected. That is why below we carry out a detailed numerical study
of the phenomenon.

\section{Simulation}

The simulation program solves dynamics of the system governed by
GPE, and is based on Split Step Fourier and the Interaction Pictures
Runge-Kutta-4 method.\cite{Ballagh}. The initial conditions are
chosen as follows. Since the macroscopic wave function is uniform
throughout the elongated trap with a tunneling tail at the barrier,
it was modeled as
\begin{equation}
\psi(x,0) = \frac{1 - \tanh{\frac{x}{a}}}{b}
\end{equation}
where $a$ and $b$ are parameters defining the slope of decay at the
barrier, and the constant height inside the trap, respectively. The
external potential barrier at the edge of the trap is taken as
\begin{equation}
U_{ext}(x)=\frac{U_0}{\cosh^2\frac{(x+\alpha)}{\beta}}
\end{equation}
where $U_0$, $\alpha$ and $\beta$ are parameters that determine the
height, shift from zero and width of the barrier, respectively. We
consider weakly interacting BEC such that the interaction parameter
$\frac{U_{int}|\psi|^2}{U_0} <1 $ in the barrier region. This
condition has to hold during the entire dynamical evolution, so that
GPE remains valid.

\subsection{Right hand side of the barrier - Emergence of blip and splitting}

\begin{figure}[here!!]
\begin{center}
\includegraphics[width=2.5cm]{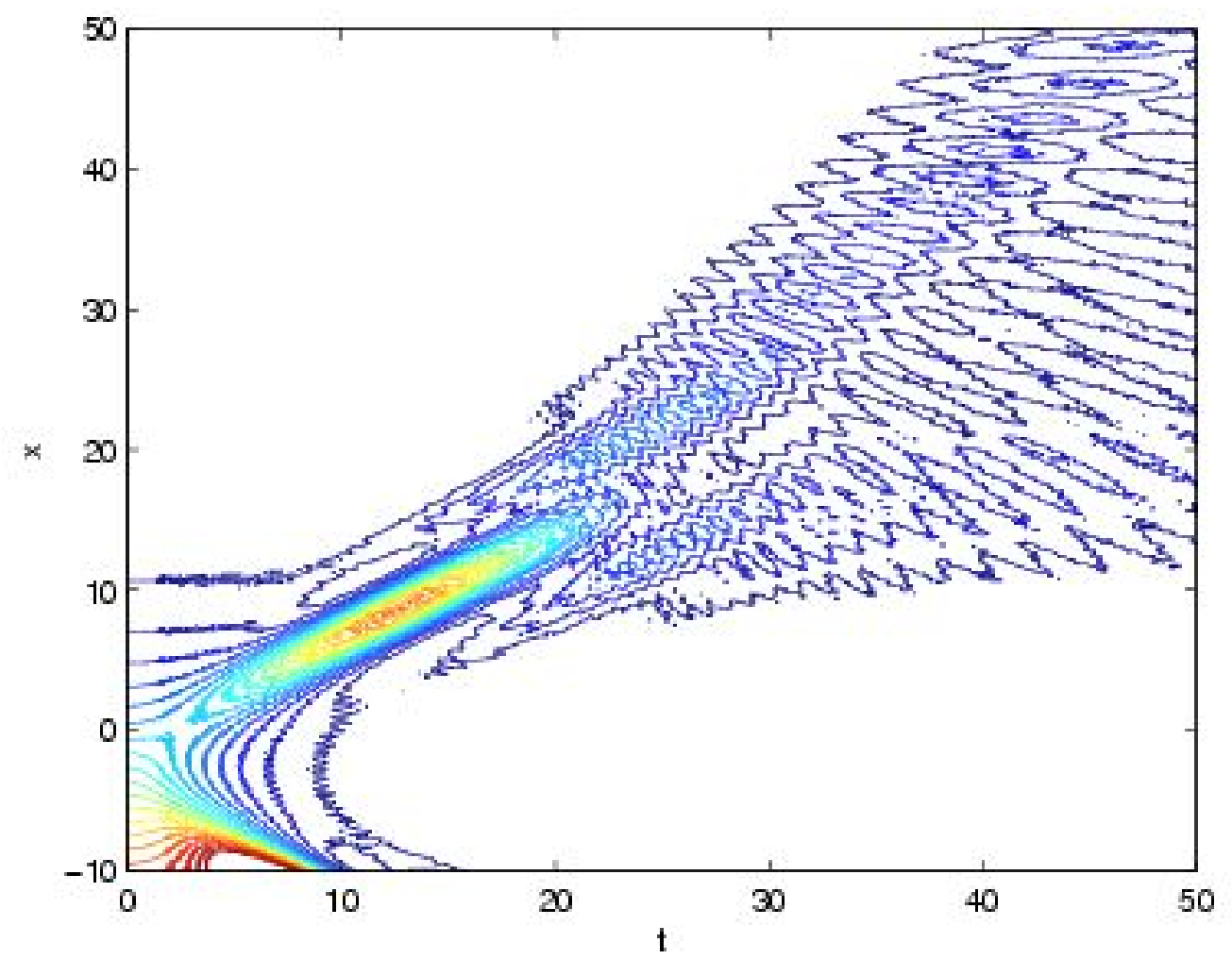}
\includegraphics[width=2.5cm]{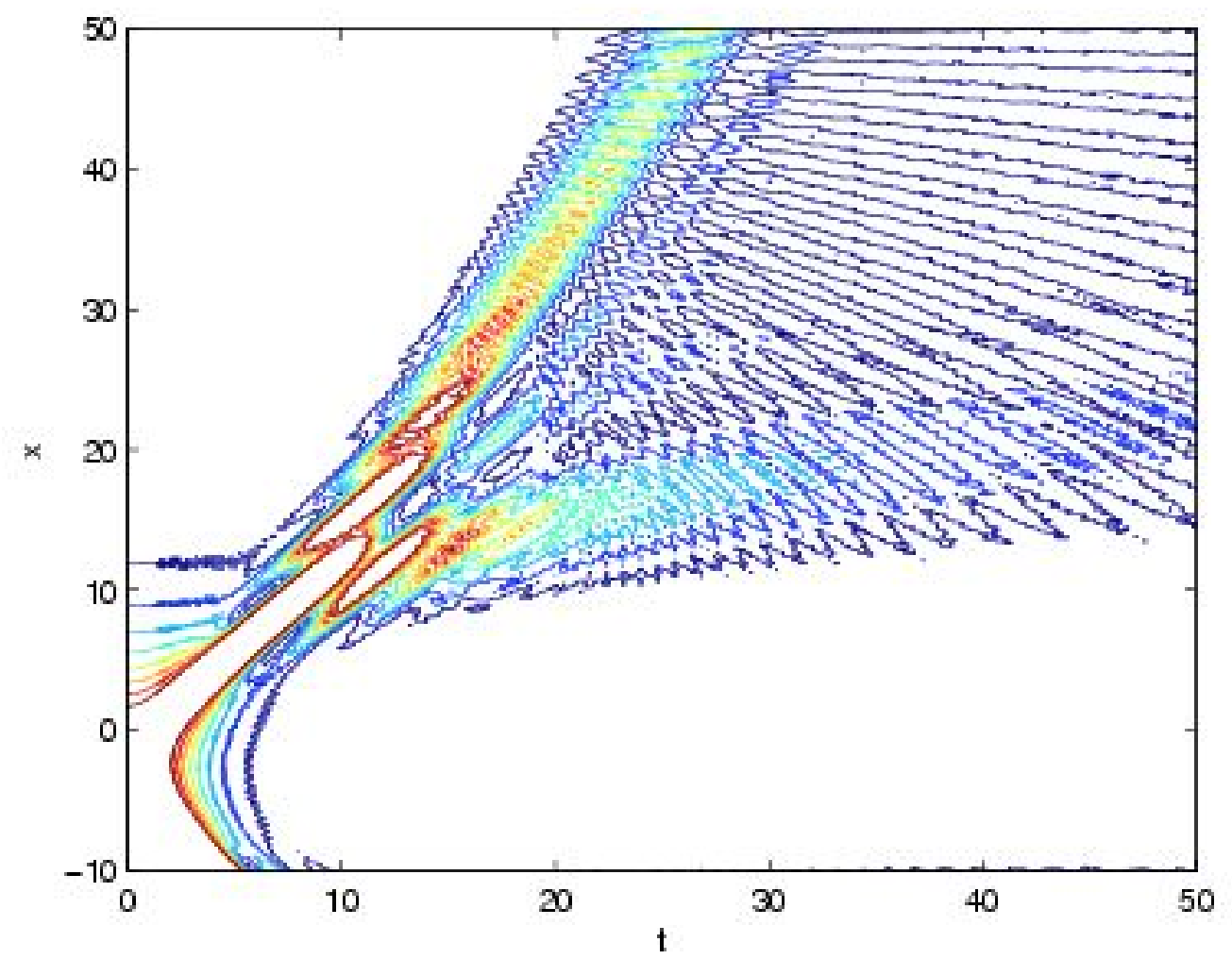}
\includegraphics[width=2.5cm]{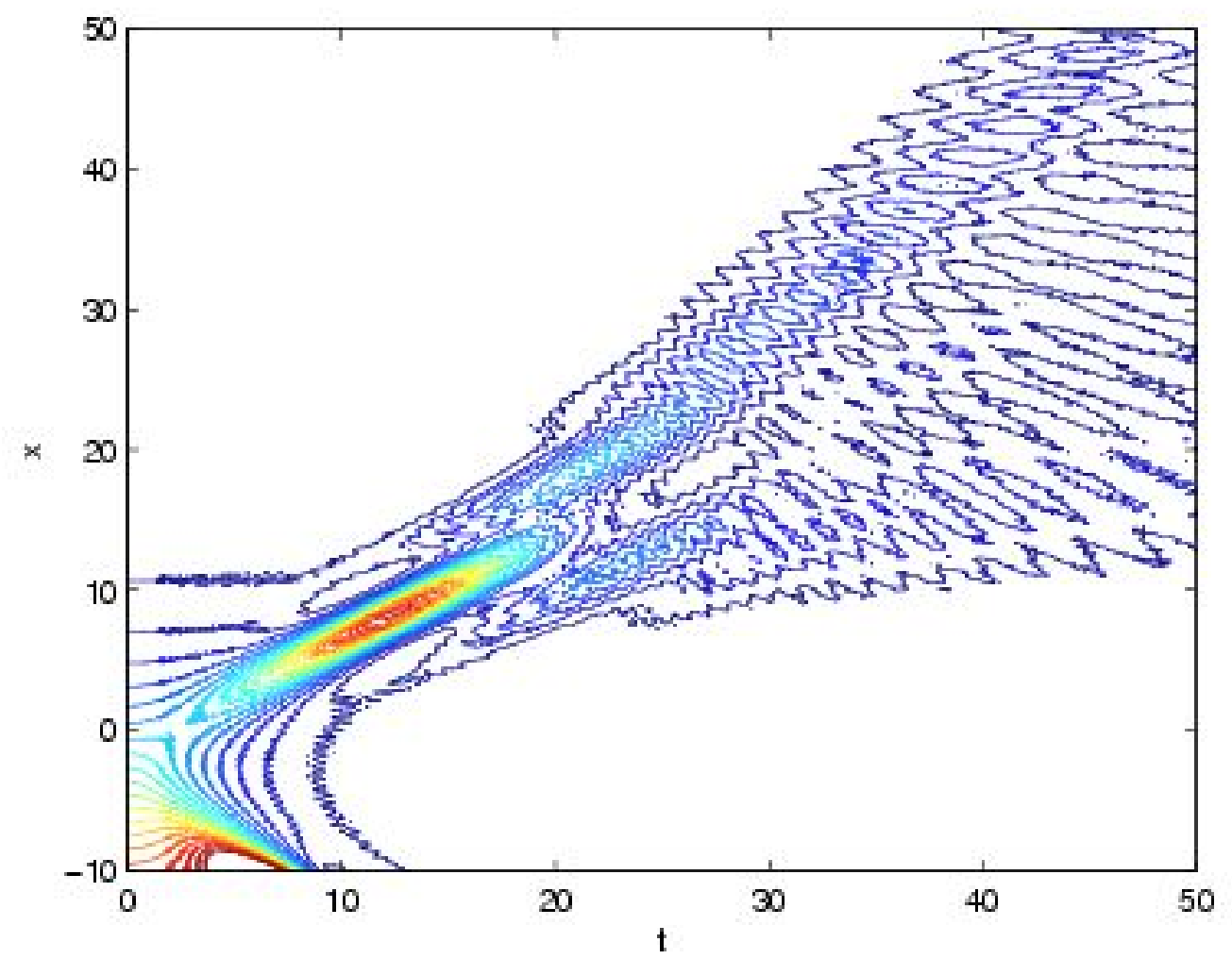}
\includegraphics[width=2.5cm]{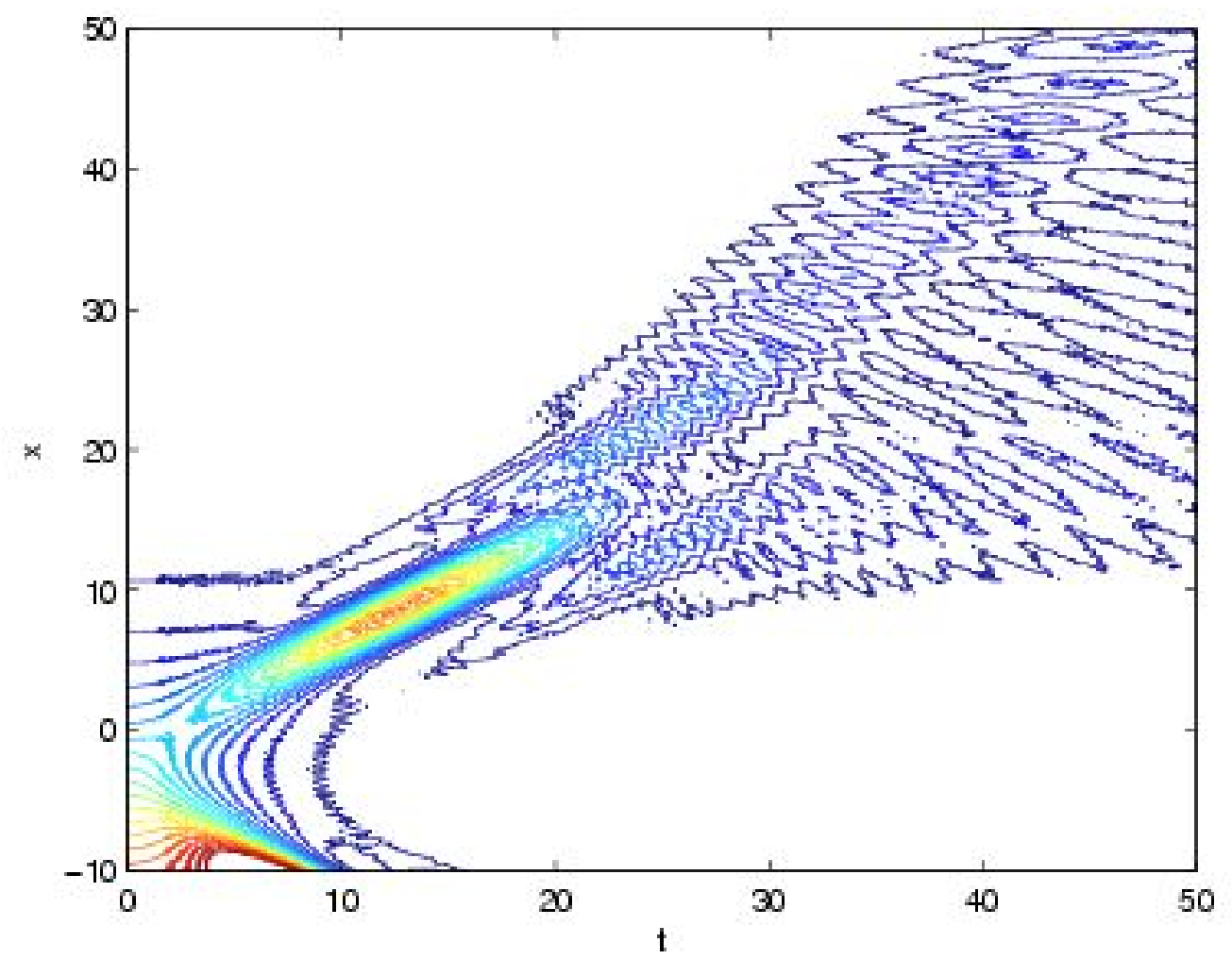}
\includegraphics[width=2.5cm]{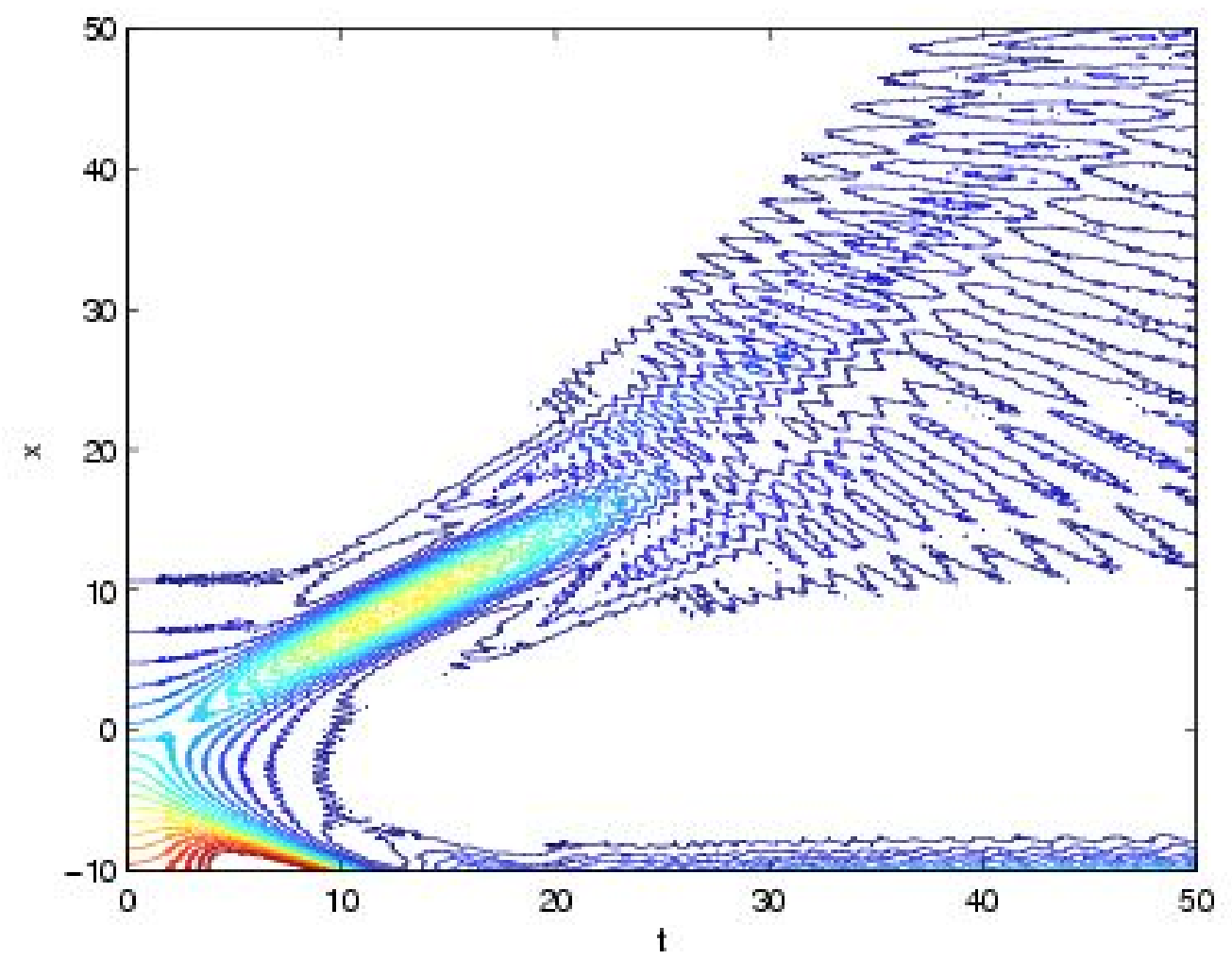}
\includegraphics[width=2.5cm]{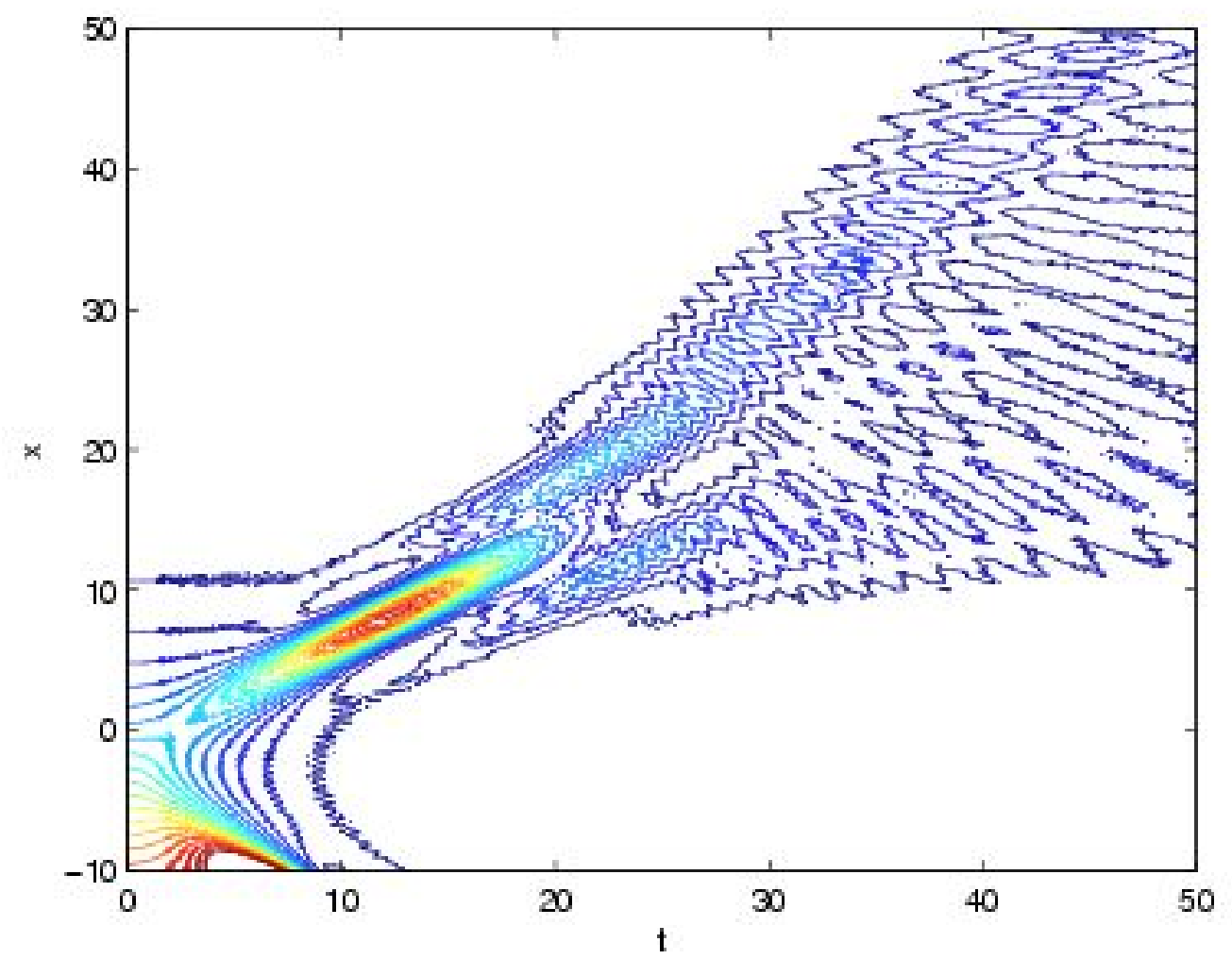}
\caption{(Color online) 1-d space coordinate vs. time plot of
numerical results showing emergence of the blip in the course of
macroscopic tunneling from a cigar shaped trap.  Upper figures
relate to barrier height of $U_0=3$, whereas in lower figures
$U_0=1$. Other trap parameters: $\alpha=3,\beta=4$. From right to
left respectively: no, repulsive($U_{int}=3$) and
attractive($U_{int}=-3$) interaction. Unlike in formerly
investigates cases, i.e. tunneling from a narrow trap, a splitting
of the blip into two components occurs right after its
emergence.}\label{blip}
\end{center}
\end{figure}

As predicted by the analytical calculation, a blip emerges for all
types of small interaction (repulsive, zero, and negative). Its
initial velocity is strictly derived from the height of the barrier
as $v = \sqrt{2U_0/m}$, also in the case of cigar shaped BEC as seen
in Fig.\ref{blip} However an interesting phenomenon arises, known to
appear when super-Gaussian states are involved. Ref.
\onlinecite{gfvg08} shows using non-linear geometric optics method
that an initial optical pulse of a super-Gaussian profile and high
enough intensity, will eventually split in time into two shorter
pulses. In accordance to the latter, splitting of the blip into two
smaller blips under certain geometric parameters is obtained in the
present work as well, where the split fractions propagate with
velocities which roughly maintain momentum conservation of inelastic
collision (if one ignores radiation): the mass of the original blip
times its velocity equals the sum of masses times velocities of the
split smaller blips. This is also seen in Fig. \ref{blip}

\subsection{Inside the barrier - depletion}\label{dep}

At the time the blip is created it leaves behind a depleted region,
as was also shown in Refs. \onlinecite{ours1,ours2}. This depletion
enables consideration of the original problem as two separate
problems in time and space, outside and inside the trap. The main
difference between the previous and the present work is that while
in the former this depletion was used to estimate the initial blip
mass, in the latter it also indicates a compensating local rise of
density in the trapped BEC near the barrier, which may ignite a
shock propagation. If the symmetry had been complete there would
have been no local increase of density inside the BEC but only an
anti-blip that would have propagated as a dark soliton. Another
interesting question is how long this depletion remains 'locked'
inside the barrier, and when does a tunneling tail starts to
refilling it. The answer to this will be discussed below.

\subsection{Left hand side of the barrier - Anti-blip?}

\begin{figure}[tbp!]
\begin{center}
\includegraphics[width=5cm]{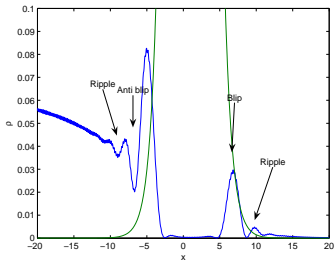}
\caption{(Color online) An almost symmetrical blip - anti-blip
solution appearing for the proper choice of trap parameters:
$U_0=3,\alpha=1,\beta=2$ for small repulsive
interaction($U_{int}=3$), at $t=10$.}\label{antiblip}
\end{center}
\end{figure}

For repulsive interaction, an interesting case arises for which the
tunneling process has almost perfect symmetry between the right and
left sides of the barrier. This was discussed previously, in
Sections \ref{left} and \ref{dep}. In the former it was shown that
if a local minimum of density inside the BEC is small enough, there
is symmetry between the right and left side of the barrier. In this
case one expects to see an anti-blip and an 'anti-ripple' inside the
BEC which propagate with the blip velocity, and which might even
become a propagating dark soliton. This, of course, has to do with
the proper geometry, and indeed seen in Fig.\ref{antiblip} where an
almost perfect symmetry between the propagating blip and anti-blip
at $t=10$ is presented.

\subsection{Left hand side of the barrier: dispersive shock wave?}

An interesting observation is the dynamical evolution inside the
BEC, which can be assigned to dispersive shock propagation in the
systems controlled by GPE. It starts with depletion inside the
barrier which remains locked until the BEC recovers, and a jump in
the particle density inside the BEC to compensate for the depletion.
This jump propagates accompanied, as is typical for dispersive
shocks, by strong oscillations in its front, and leaving a slower
rarefication behind. An important fact is that this is not sound
propagation, as these shock waves evolve and propagate also in the
limit of zero interaction when the sound velocity becomes zero. Let
us examine the two other cases.

\begin{figure}[tbp!]
\begin{center}
\includegraphics[width=4cm]{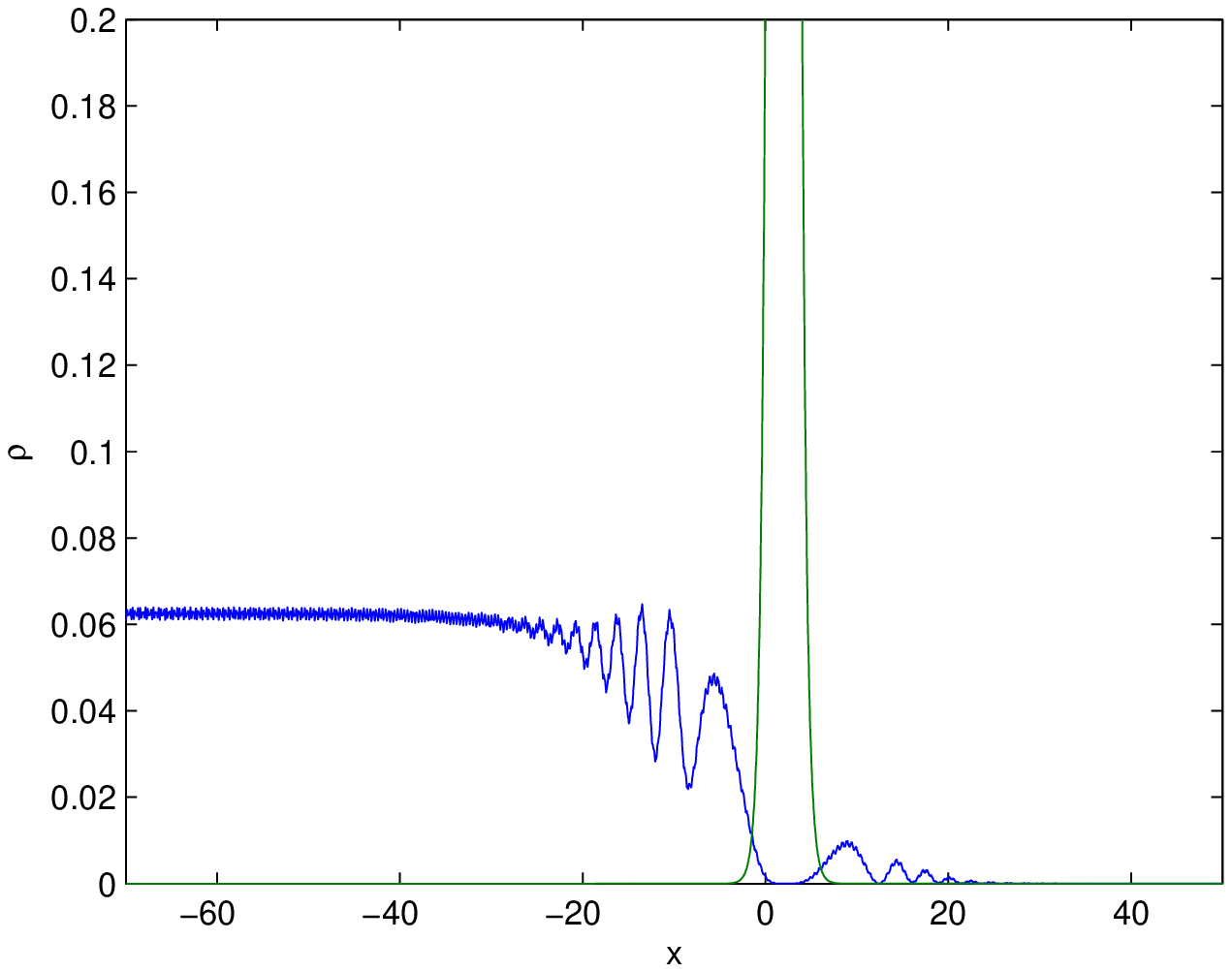}
\includegraphics[width=4cm]{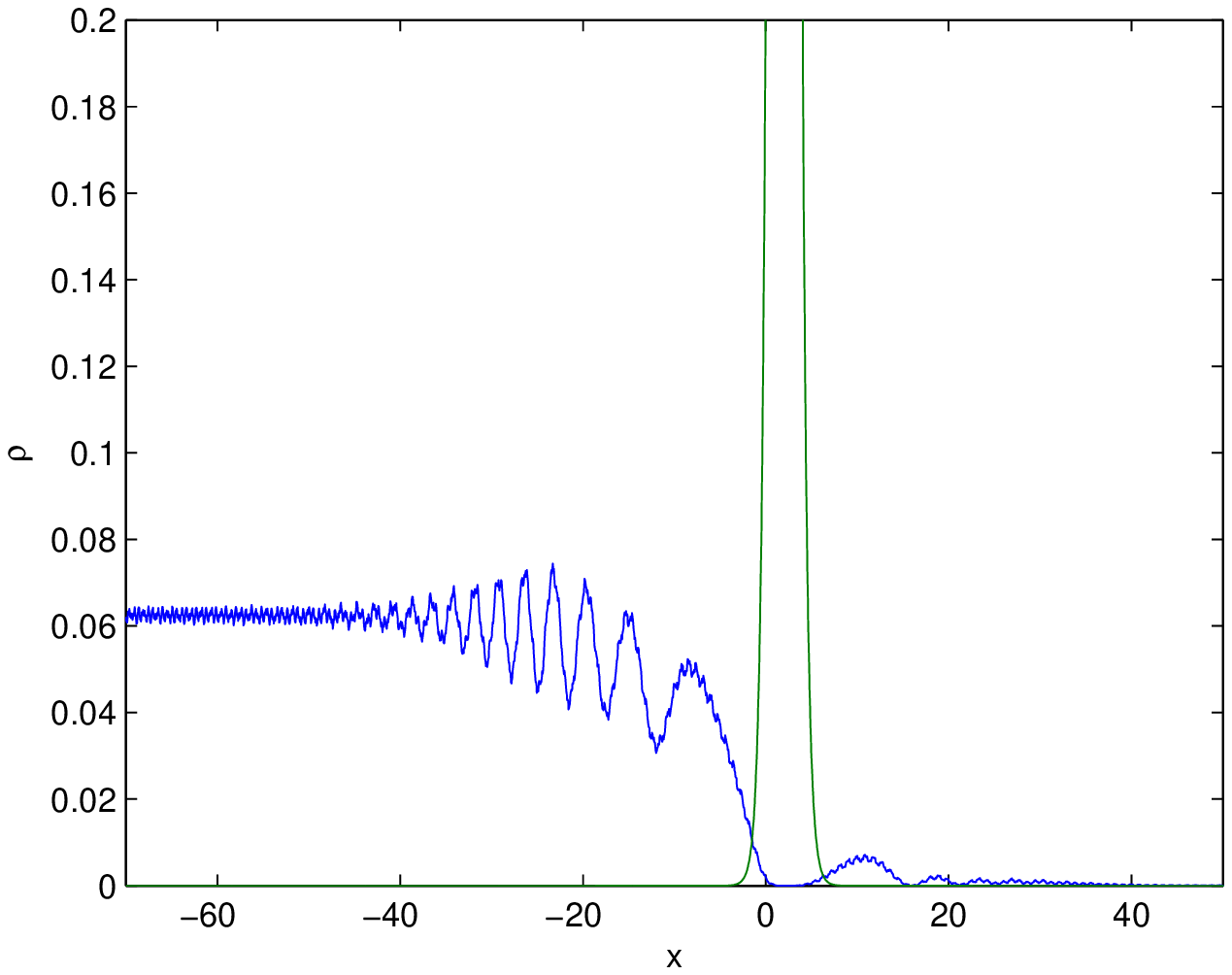}
\includegraphics[width=4cm]{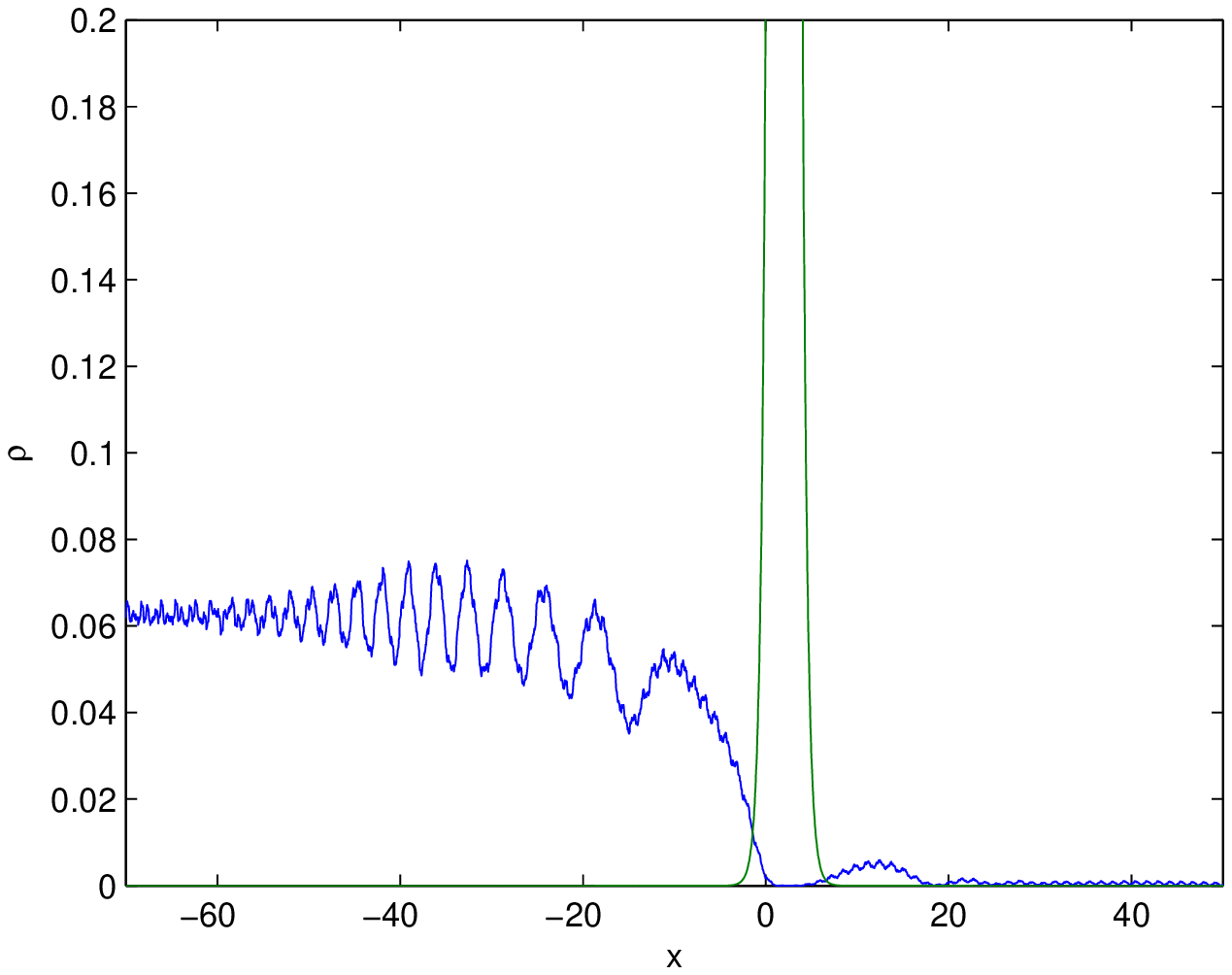}
\includegraphics[width=4cm]{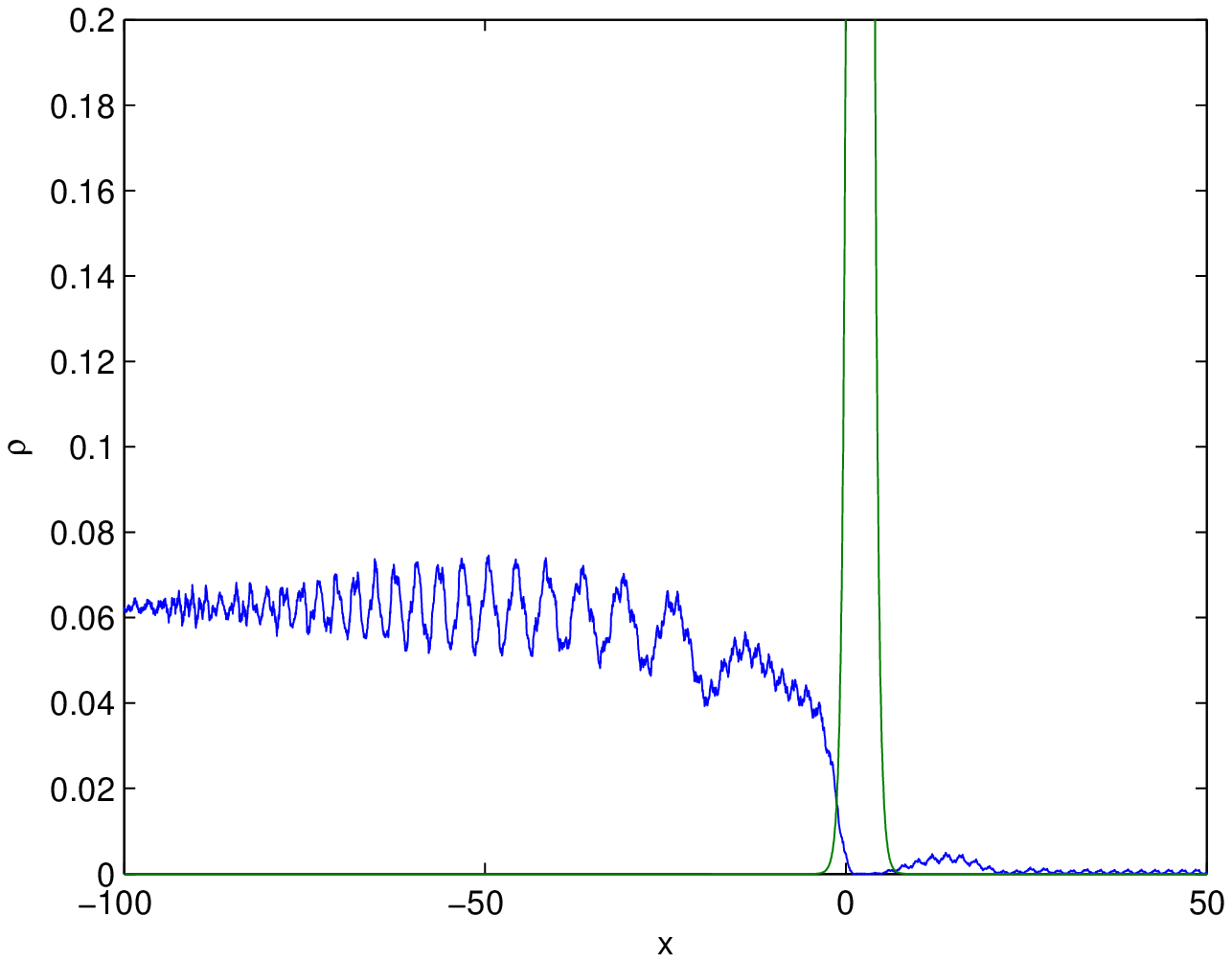}
\caption{(Color online) Particle density vs. 1-d space coordinate
plots show propagation of a jump in the particle density inside the
BEC induced by tunneling for repulsive interaction. Increasing times
from upper left figure are: $t=20,70,100,150$. This matches the
dynamic behavior of dispersive shock propagation. In the course of
the evolution it can be seen that the packet recovers to a new
constant amplitude near the barrier, after it lost the blip mass.}
\label{repulsive}
\end{center}
\end{figure}

\paragraph{Repulsive interaction.}
The dynamical solution inside the BEC for repulsive interaction is
presented in Fig. \ref{repulsive}. In this case a dispersive shock
propagation is clearly observed. An important note is that the
velocity of its front matches that of the blip, traveling on the
right side of the barrier, and unlike in normal sound waves has no
connection to the interaction strength. After long enough time the
BEC exhibits a new equilibrium height, that might enable repeatable
processes, Fig. \ref{repulsive}.

\begin{figure}[tbp!]
\begin{center}
\includegraphics[width=5cm]{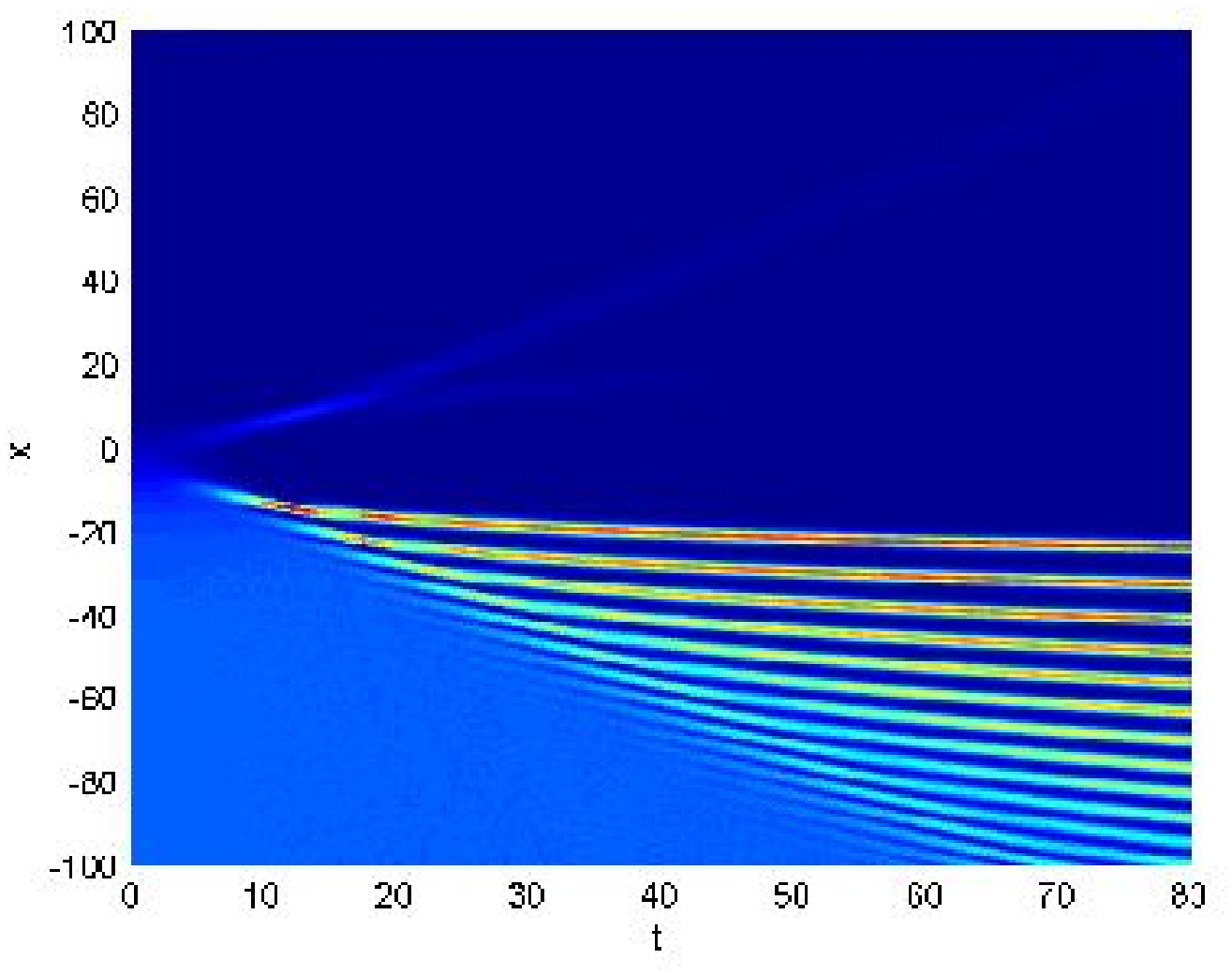}
\includegraphics[width=5cm]{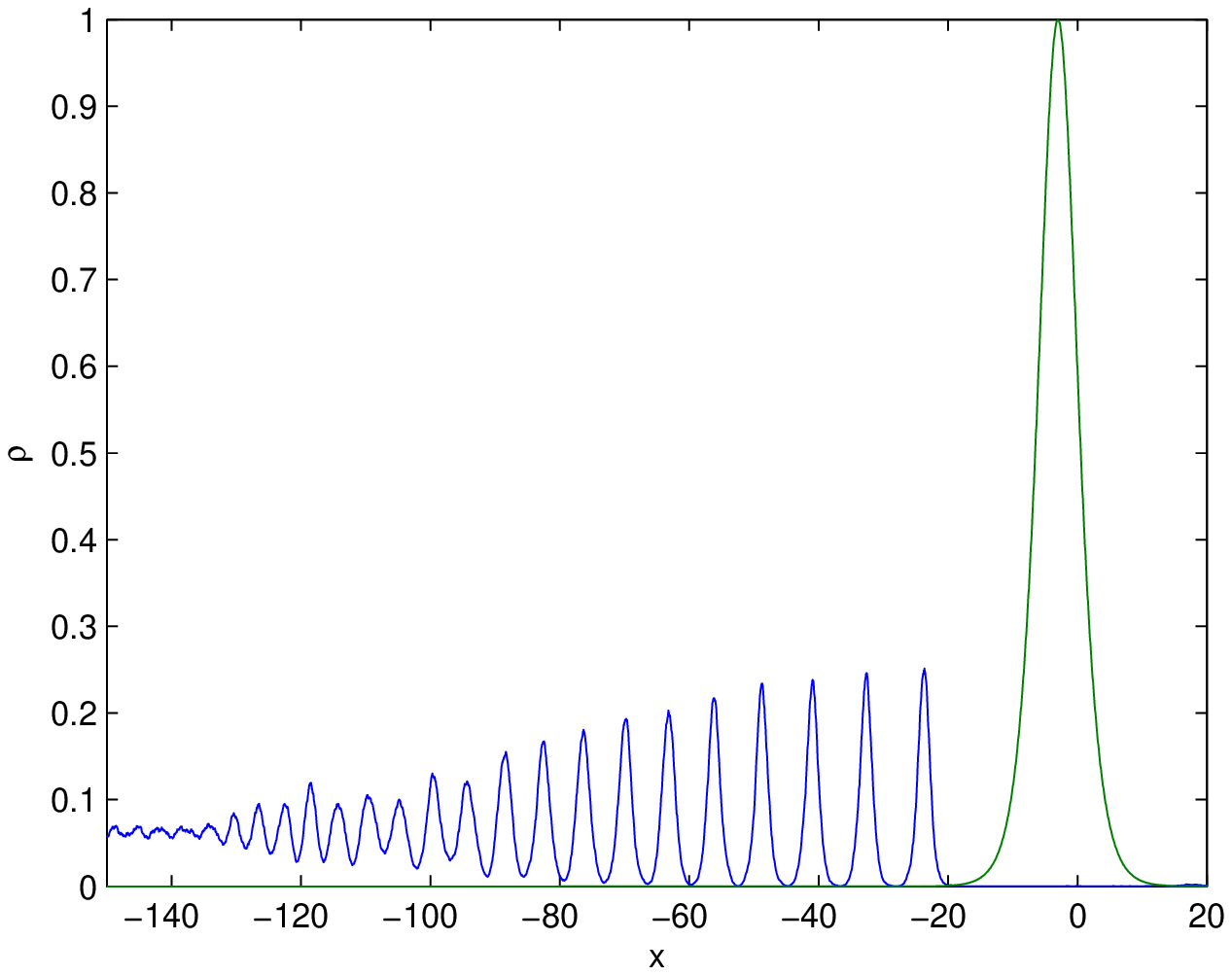}
\caption{(Color online) Inside the BEC - attractive interaction
case. A shock does not propagate, instead, localized bright solitons
are formed. This is seen in both upper figure which shows 1-d space
coordinate vs. time plot and lower figure which is a frozen density
vs. 1-d space coordinate plot at $t=200$.} \label{attractive}
\end{center}
\end{figure}

\paragraph{Attractive interaction- localized train of bright solitons.}
For attractive interaction a shock wave is eventually damped and
does not propagate. Parts of the BEC become localized bright
solitons shown in Fig. \ref{attractive}.

\section{Conclusions}

In this work we have investigated the dynamics of macroscopic
tunneling from an elongated cigar shaped trap. Several interesting
phenomena were predicted by analytic formalisms and by numerics. As
in the previous works\cite{ours1,ours2} on macroscopic tunneling, a
blip in the particle density was shown to appear in the margins of
the potential barrier and propagate with constant velocity, but
contrary to other configurations, it was shown to split into two
quite fast. Inside the BEC, dispersive shock waves that are now
known to emerge in numerous configurations in BEC setups with
repulsive interactions were predicted to appear, this time in
macroscopic tunneling problems, and their velocity was shown to be
easily controlled. In the case of attractive interaction bright
solitons were shown to localize near the barrier. The BEC was also
shown to stabilize near the barrier, which might enable realization
of a soliton lasing from such a system. Symmetry and antisymmetry
between the left and right hand sides of the barrier were discussed,
and exhibited for a proper choice of parameters in a propagating
'anti- blip' that might become a 'gray soliton'.

{\bf Acknowledgments.} The authors acknowledge the support of United
States - Israel Binational Science Foundation, Grant N 2006242. A.S.
is partially supported by NSF. Hospitality of Max Planck Institute
for Complex Systems, Dresden is highly appreciated.

\end{document}